\documentclass[reprint,
bibnotes,
amsmath,amssymb,
aps,
]{revtex4-2}
\usepackage[justification=centerlast]{caption}
\usepackage{subcaption}
\usepackage{hyperref}
\usepackage{indentfirst}
\usepackage{amsmath}  
\usepackage{graphicx}
\usepackage{dcolumn}
\usepackage{float}
\usepackage{bm}
\usepackage{xcolor}
\usepackage{comment}

\begin{document}
\title{General solution for the response of materials under radiation and tilted magnetic field: semi-classical regime}

\author{Narjes Kheirabadi$^{1, 2}$, YuanDong Wang$^{3,4,5}$}%
 
\affiliation{$^1$ Department of Physics, Iran University of Science and Technology, Narmak, Tehran 16844, Iran\\
$^2$ Department of Physics, University of Tehran, Tehran, 14395547, Iran
}
\affiliation{
$^3$ Department of Applied Physics, College of Science, China Agricultural University, Qinghua East Road, Beijing 100083, China\\
$^4$ School of Electronic, Electrical and Communication Engineering, University of Chinese Academy of Sciences, Beijing 100049, China\\
$^5$ School of Physical Sciences, University of Chinese Academy of Sciences, Beijing 100049, China}


\begin{abstract}
The Berry curvature dipole is well-known to cause Hall conductivity. This study expands on previous results to demonstrate how two- and three-dimensional materials react under a tilted magnetic field in the linear and nonlinear regimes. We show how the Hall effect has a quantum origin by deriving the general form of intrinsic and extrinsic currents in materials under a tilted magnetic field. Our focus is on determining the linear and nonlinear response of two-dimensional materials. We also demonstrate that as the result of the perpendicular component of the magnetic field a current resulted by both velocity and Berry curvature can occur in two-dimensional materials and topological crystalline insulators in second harmonic generation and ratchet responses. The findings of this research may provide insight into the transport characteristics of materials in the semi-classical regime and initiate a new chapter in linear and nonlinear Hall effects.
\end{abstract}

\maketitle
\section{Introduction}
Electron wave velocity is the primary current source in a material under radiation. There are two sources for the electron wave velocity: classical velocity, which comes from the energy dispersion form, and anomalous velocity, which comes from the Berry curvature (BC) \cite{sundaram1999wave, xiao2010berry}. The BC is a physical parameter that could be determined by the eigenstates and eigenvalues for a specific Hamiltonian, so it is completely related to the quantum mechanical characteristic of a material \cite{xiao2010berry}. 
It is not fundamental to break the time-reversal symmetry to have the nonlinear Hall effect, but breaking inversion symmetry seems to result in a nonlinear Hall effect. Subsequently, the nonlinear Hall effect could be a way to investigate the symmetry and the topology of materials' quantum phases. When the crystal symmetry is the origin of the nonlinear Hall effect, some unconventional responses such as the nonlinear spin Hall effect \cite{hamamoto2017nonlinear, araki2018strain}, the Magnus Hall effect \cite{papaj2019magnus}, the gyrotropic Hall effect \cite{konig2019gyrotropic}, and the nonlinear Nernst effect \cite{yu2019topological, zeng2019nonlinear} are also possible (for a review on these effects see Ref.~\cite{du2021nonlinear}).

While extrinsic responses are dependent on the scattering rate of electrons, the intrinsic transport responses are independent of the scattering time, and these types of responses excitingly reveal the geometry of the band structure \cite{gao2014field, liu2021intrinsic, PhysRevLett.127.277201, huang2023intrinsic, wang2024orbital}. In experiments, it is possible to separate intrinsic responses from scattering-dependent responses, as well \cite{kang2019nonlinear, gao2023quantum,wang2023quantum}. It has been shown that intrinsic currents originate from Berry-connection polarizability (BCP) and its spin susceptibility \cite{huang2023intrinsic}, the orbital magnetic moment  \cite{das2021intrinsic, huang2023nonlinear}, and also BC correction under the applied magnetic field \cite{wang2024orbital}. Also, nonlinear responses of spin-orbit coupled 2-dimensional (2D) materials under a parallel magnetic field and in the DC limit where the distribution function of electrons is time-independent have been studied recently \cite{huang2023nonlinear}. 

In this article, we have extended previous results and we have found the contribution of a steady tilted magnetic field in the deduced current for materials under an AC electric field. After that, our focus was on 2D materials and a potential candidate, topological crystalline insulators (TCIs). All linear and nonlinear quantum and classical currents in materials were generated when time-reversal symmetry was broken by a tiled magnetic field.  According to our findings, materials exhibit a linear intrinsic Hall current that can be predicted based on the BC and its correction under the applied magnetic field. In TCIs, a linear intrinsic current proportional to the perpendicular magnetic field can be predicted, as we have shown.  We have also derived a current resulted by both velocity and BC nonlinear Hall conductivity,  which is solely dependent on the application of a perpendicular magnetic field. This nonlinear response is dependent on the second derivative of the multiplication of classic velocity and BC. The results of this work are valid for all radiation polarizations of THz and microwave types where the semi-classical regime is valid and the field-induced electron energy \cite{gao2014field}, inversion asymmetric scattering rate, and the side-jump mechanism are ignored \cite{olbrich2014room, deyo2009semiclassical, du2021nonlinear}. This work is also valid when the magnetic field is relatively weak so that the mean free path is less than the cyclotron radius. So, the Bohr-Sommerfeld quantization and Landau levels cannot be observed \cite{chen2015hofstadter, kheirabadi2018cyclotron}.
\section{Boltzmann kinetic equation}
Under radiation, the electron distribution function is out of equilibrium, and the velocity of the electron wave is determined by the classical velocity and the Berry curvature, $\mathbf{\Omega}$, which originates anomalous velocity \cite{sodemann2015quantum}. Under the external electric and magnetic fields, $\mathbf{E}$ and $\mathbf{B}$, the field-corrected Berry curvature $\tilde{\mathbf{\Omega}}$ is the sum of $\mathbf{\Omega}$ and $\mathbf{\Omega}^\prime$, where $\mathbf{\Omega}^{\prime}=\mathbf{\nabla}\times\bm{\mathcal{A}}^\prime$, $\bm{\mathcal{A}}^\prime = \bm{\mathcal{A}}^\prime (E) +\bm{\mathcal{A}}^\prime (B)$. For the $nth$ band, we have
\begin{equation}\label{Eq1}
\bm{\mathcal{A}}^{\prime}_{n}(E)=\overleftrightarrow{\bm{G}}_{n}\bm{E}, \quad \bm{\mathcal{A}}^{\prime}_{n}(B)=(\overleftrightarrow{\bm{F}}_{n}^{S}+\overleftrightarrow{\bm{F}}_{n}^{O})\bm{B},
\end{equation} 
where the second rank tensors $\overleftrightarrow{\bm{G}}$, the Berry connection polarizability (BCP), which is written as \cite{gao2014field,gao2019semiclassical,wang2023field}
\begin{eqnarray}\label{Gtensor}
G_n^{\alpha \beta}=2 \sum_{m\neq n} \frac{ \text{Re}[A_{nm}^{\alpha} A_{mn}^{\beta}]}{\varepsilon_n- \varepsilon_m},
\end{eqnarray}
$\overleftrightarrow{\bm{F}}_{n}^{S}$ and $\overleftrightarrow{\bm{F}}_{n}^{O}$ are the anomalous spin polarizability (ASP) and anomalous orbital polarizability (AOP), which are given by \cite{gao2014field, PhysRevB.103.045401}
%
\begin{equation}\label{FtensorS}
F^{S,\alpha\beta}_{n}=-2\text{Re}\sum_{m\neq n}\frac{\mathcal{M}_{nm}^{S,\alpha}\mathcal{A}^{\beta}_{mn}}{\varepsilon_n -\varepsilon_m},
\end{equation}
\begin{equation}\label{FtensorO}
F^{O,\alpha\beta}_{n}=-2\text{Re}\sum_{m\neq n}\frac{\mathcal{M}_{nm}^{O,\alpha}\mathcal{A}^{\beta}_{mn}}{\varepsilon_n -\varepsilon_m}-\frac{1}{2}\epsilon^{\alpha\gamma\zeta}\partial^{\gamma}\mathcal{G}_{n}^{\alpha\beta}.
\end{equation}
Here, $A_{nm}=i \left\langle u_{n \mathbf{k}}\left| \nabla _{\mathbf{k}} \right| u_{m\mathbf{k}} \right\rangle$ is the interband Berry connection, $\varepsilon_n$ is unperturbed band energy. $\bm{\mathcal{M}}^{S}_{mn}=-g\mu_B \bm{S}_{mn}$ and $\bm{\mathcal{M}}^{O}_{mn}=\sum_{l\neq n}(\bm{v}_{ml}+\delta_{lm}\bm{v}_{n})\times \bm{\mathcal{A}}_{lm}/2$ are the interband spin and orbital magnetic moments, respectively. Here, $g$ is the g-factor for spin, $\mu_B$ is the Bohr magneton, and $\bm{S}_{mn}$ is the matrix elements of spin operator, $\epsilon^{\alpha\gamma\zeta}$ is the Levi-Civita symbol, and $\mathcal{G}_{n}^{\alpha\beta}=\text{Re}\sum_{m\neq n}\mathcal{A}_{nm}^{\alpha}\mathcal{A}_{mn}^{\beta}$ is known as the quantum metric tensor.
We also have $\bm{v}_n$, the group velocity of the electron wave, is $\partial \varepsilon_n/\partial \mathbf{k}$ ($\hbar=1$), so we assume that $\tilde{\mathbf{\Omega}}=\mathbf{\Omega}+\mathbf{\Omega}^E+\mathbf{\Omega}^B$.

Additionally, the equations of motion are \cite{xiao2010berry} 
\begin{eqnarray}\label{rdot}
D(\mathbf{B},\tilde{\mathbf{\Omega}})\mathbf{V}&&=\mathbf{V_{cl}}+e \mathbf{E}\times \tilde{\mathbf{\Omega}}+e(\mathbf{V_{cl}}\cdot \tilde{\mathbf{\Omega}}) \mathbf{B},\\
D(\mathbf{B},\tilde{\mathbf{\Omega}})\dot{\mathbf{k}}&&= -e \mathbf{E}-e \mathbf{V_{cl}} \times \mathbf{B} - e^2 (\mathbf{E} \cdot \mathbf{B}) \tilde{\mathbf{\Omega}}.
\end{eqnarray}
in which, the phase space factor, $D(\mathbf{B},\tilde{\mathbf{\Omega}})=1+ e \bm{B}\cdot \tilde{\bm{\Omega}}$, $e>0$ is the electron charge, $\mathbf{k}$ is the electron wave vector, $\mathbf{V}_{cl}$ is the classical velocity of the electron, and we consider the responses up to the linear order in $B$.
We can show that according to the above assumptions, we have
\begin{equation}\label{kdot}
\begin{aligned}
\dot{\mathbf{k}}=-e \mathbf{E}-e \mathbf{V_{cl}} \times \mathbf{B} + e^2 \mathbf{B} \times (\mathbf{E} \times (\bm{\Omega}+\bm{\Omega}^E)).
\end{aligned}
\end{equation}

In the relaxation time approximation, the Boltzmann kinetic equation (BKE) is \cite{mahan2013many}
\begin{eqnarray}\label{bke}
\frac{\partial f}{\partial t} + \dot{\mathbf{k}} \cdot \nabla f=-\frac{f-f_0}{\tau}. 
\end{eqnarray}
where $f$ is the nonequilibrium distribution function of the electron at a time, $t$, and the scattering time is $\tau$.
We also assume that the nonequilibrium distribution function of the electron obeys the following form \cite{sodemann2015quantum}
\begin{eqnarray}\label{f0}
f=Re\bigg[f_0^0+f_2^0+f_1 e^{i \omega t}+f_2 e^{2 i \omega t} \bigg].
\end{eqnarray}
By the substitution of Eqs. \ref{kdot} and \ref{f0} in Eq. \ref{bke}, and by multiplying the Boltzmann kinetic equation in $1/T\int_{0}^{T} e^{iL \omega T} dt$ where $L$ is an integer and the integrate is over a period, T, coupled equations between different harmonic coefficients are achieved. In addition, considering those terms linear in the magnetic field and doing the calculations up to the second order in the electric field (an applied electric field for radiation with the angular frequency  $\omega$ is $\mathbf{E} = 1/2 \big(\mathbf{\mathcal{E}}\exp{(i \omega t)}+\mathbf{\mathcal{E}}^*\exp{(-i \omega t)}\big)$), we can show that different harmonics are

\begin{equation}\label{f1}
f_1 = \left(  \mathcal{W}^{a}+  \mathcal{W}^{ia}_{B} B_{i}  +\mathcal{W}^{bga}_{B,\Omega} B_b\Omega_{g} 
\right)\mathcal{E}_a.
\end{equation}
Here we introduce the coefficient tensors $\mathcal{W}$ which are classified by the dependences on $\bm{B}$ and $\bm{\Omega}$: 
\begin{equation}
\begin{aligned}
&\mathcal{W}^{a}=\frac{ \tau e}{1 + i \omega \tau}\partial_a f_0^0,\\
&\mathcal{W}^{ia}_{B}=\tau^2 e^2\frac{2+i\omega \tau}{(1+i\omega\tau)^2}  \varepsilon_{ghi} V_{cl_{h}}\partial_{ag} f_0^0, \\
&\mathcal{W}^{bga}_{B,\Omega}=-\frac{ \tau e^2}{1 + i \omega \tau} \varepsilon_{bcd}\varepsilon_{agc} \partial_d f_0^0. \\
\end{aligned}
\end{equation}
where $\varepsilon_{abc}$ denotes the Levi-Civita symbol. 
\begin{equation}
\begin{aligned}
f_2 =&\left(\mathcal{V}^{ja}\mathcal{E}_j + \mathcal{V}^{bgda}_{B,\Omega}B_b\Omega_{g}\mathcal{E}_d   + \mathcal{V}^{ika}_{B} B_i\mathcal{E}_k  \right)\mathcal{E}_a\\
&+ \mathcal{V}^{bdg}_{B,P}P_{da}B_{b}\mathcal{E}_g\mathcal{E}_a,
\end{aligned}
\end{equation}
with the coefficient tensors given by
\begin{equation}
\begin{aligned}
&\mathcal{V}^{ja}=\frac{ \tau^2 e^2}{2(1 + i \omega \tau)(1 + 2 i \omega \tau)}\partial_{ja} f_0^0,\\
&\mathcal{V}^{bgda}_{B,\Omega}=- \frac{ \tau^2 e^3}{(1 + i \omega \tau)(1 + 2 i \omega \tau)} \varepsilon_{bcj}\varepsilon_{dgc} \partial_{ja} f_0^0, \\
&\mathcal{V}^{ika}_{B}=-  \tau^3 e^3 \frac{ 1+4i\omega\tau -2\omega^2\tau^2}{2(1 + i \omega \tau)^2 (1+2i\omega\tau)^2}  \varepsilon_{ghi} V_{cl_h}  \partial_{akg}f_0^0, \\
&\mathcal{V}^{bdg}_{B,P}=\frac{\tau e^2}{2 (1+2i\omega \tau)}  \varepsilon_{bcl} \varepsilon_{dgc}     \partial_l f_0^0, \\
\end{aligned}
\end{equation}
in which we define a Berry curvature polarizability as $P_{da}=\varepsilon_{dhk}\partial_h G_{ka}$ \cite{PhysRevB.105.045118}. We also have
\begin{equation}\label{f20}
\begin{aligned}
f_2^0=&  (  \mathcal{\chi}^{ja}\mathcal{E}_{j}  + \mathcal{X}^{cka}_{B}B_c\mathcal{E}_k)\mathcal{E}_a^*\\
& +\mathcal{X}^{blg}_{B,P}B_b({\mathcal{E}_l}^*\mathcal{E}_g +\mathcal{E}_{l}\mathcal{E}_{g}^*)+\tau e \varepsilon_{abc} V_{cl_b}B_c \partial_a f_{0}^0, 
\end{aligned}
\end{equation}
with  
\begin{equation}
\begin{aligned}
&\mathcal{X}^{ja}=\frac{ \tau^2 e^2}{2(1 + i \omega \tau)}\partial_{ja}f_{0}^0,\\
&\mathcal{X}^{cka}_{B}=\tau^3 e^3{\frac{  (3+2i\omega\tau)}{2(1 + i \omega \tau)^2}} \varepsilon_{dbc} V_{cl_b} \partial_{dak}f_0^0,\\
&\mathcal{X}^{blg}_{B,P}= \frac{\tau e^2}{4} \varepsilon_{abc}    \varepsilon_{dgc} P_{dl}  \partial_a f_0^0.
\end{aligned}
\end{equation}
%
\section{The form of the currents (linear, ratchet, and second harmonic generation (SHG))}
The form of the current density in the $a$ direction is dependent on the nonequilibrium distribution function, the electron velocity, the electron charge, and the phase space factor according to the following equation
\begin{eqnarray}\label{j}
j_a=-e\int f(k) D(\mathbf{B},\bm{\tilde{\Omega}})V_a,
\end{eqnarray}
where $\int \equiv \int (dk)^d/(2 \pi)^d$, $d$ is the dimension of the material. To find the current density in the $a$ direction, it is enough to substitute the distribution function and the electron velocity (Eqs.~\ref{rdot} and \ref{f0} in Eq.~\ref{j}); so, we have
\begin{equation}\label{ja}
\begin{aligned}
j_a=&\text{Re}\bigg[-e\int \big(f_0^0+f_2^0+f_1 e^{i \omega t}+f_2 e^{2 i \omega t} \big)\\
&\times\big( V_{cl_a}+e(\mathbf{V}_{cl} \cdot \bm{\tilde{\Omega}}) B_a+e \varepsilon_{auv} E_u \tilde{\Omega}_v\big) \bigg].
\end{aligned}
\end{equation}

If we assume that $S=V_{cl_a}+e \big(  \mathbf{V}_{cl} \cdot \mathbf{\Omega}\big) B_a$, regarding $S \in \Re$, we can rewrite Eq.~\ref{ja} as the following form
\begin{equation}
\begin{aligned}
j_a=&\text{Re}\bigg[-e\int \big(f_0^0+f_2^0+f_1 e^{i \omega t}+f_2 e^{2 i \omega t} \big) \\
&\times\big( S+e(\mathbf{V}_{cl} \cdot \bm{\Omega}^E) B_a +e \varepsilon_{auv} E_u (\Omega_v+\Omega^E_v+\Omega^B_v)\big) \bigg].
\end{aligned}
\end{equation}
By the substitution of the $E_u$ definition in the above equation, we have
\begin{equation}\label{ja11}
\begin{aligned}
j_a=&\text{Re}\bigg[-e\int \big(f_0^0+f_2^0+f_1 e^{i \omega t}+f_2 e^{2 i \omega t} \big) \\
& \quad \quad \quad \quad \times\big( S +  P(E)+Q(E^2) \big) \bigg].
\end{aligned}
\end{equation} 
where
\begin{eqnarray}\label{p}
P(E)&&=\frac{e}{2}  (\mathcal{E}_u e^{i \omega t} +\mathcal{E}_u^* e^{-i \omega t} )\times\nonumber\\
&&(\varepsilon_{auv}(\Omega_v +\Omega_{v}^B) + \varepsilon_{cdb} V_{cl_b}B_a\partial_c G_{du} ),
\end{eqnarray}
\begin{eqnarray}\label{q}
Q(E^2)&&=\frac{e}{4}  (\mathcal{E}_u e^{i \omega t} +\mathcal{E}_u^* e^{-i \omega t} )\times\nonumber\\
&& \varepsilon_{auv} \varepsilon_{vjk}(\mathcal{E}_l e^{i \omega t} +\mathcal{E}_l^* e^{-i \omega t} )  \partial_j  G_{kl}, 
\end{eqnarray}
and the last term of Eq.~\ref{p} is nonzero for 3D materials. 

Additionally, the general form of the current is  
\begin{eqnarray}\label{ja2}
j_a=Re \bigg[ j_a^0 + j_a^1 e^{i \omega t} + j_a^2 e^{2 i \omega t} \bigg].
\end{eqnarray}
where $j_a^0$ is related to the ratchet (DC) response, $j_a^1$ denotes the linear response and $j_a^2$ is related to the SHG.
By simplification of Eq.~\ref{ja11}, we can show that 
\begin{eqnarray}\label{ja0}
j_a^0=&&-e \int f_2^0 S\nonumber\\
&& - \frac{e^2}{2} \int f_1
(\varepsilon_{auv} (\Omega_v+\Omega_{v}^B) +\varepsilon_{cdb} V_{cl_b}B_a\partial_c G_{du}) \mathcal{E}_u^*\nonumber\\
&&-\frac{e^2}{4}\int \big(f_{0}^0+ \frac{f_2^0}{2} \big) (\mathcal{E}_u \mathcal{E}_{l}^* +\mathcal{E}_{u}^*\mathcal{E}_{l})\varepsilon_{auv} \varepsilon_{vjk} \partial_j G_{kl}, 
\end{eqnarray}
\begin{eqnarray}\label{ja2}
j_a^2=&&-e \int f_2 S \nonumber\\
&&- \frac{e^2}{2} \int f_1 (\varepsilon_{auv}(\Omega_v +\Omega_{v}^B) + \varepsilon_{cdb} V_{cl_b}B_a\partial_c G_{du} ) \mathcal{E}_u\nonumber\\
&&-\frac{e^2}{4}\int f_{0}^0 (\mathcal{E}_u \mathcal{E}_{l} +\mathcal{E}_{u}^*\mathcal{E}_{l}^*)\varepsilon_{auv} \varepsilon_{vjk} \partial_j G_{kl} , 
\end{eqnarray}
\begin{eqnarray}\label{ja1}
j_a^1=&&-e \int f_1 S \nonumber\\
&&- e^2 \int f_0^0 (\varepsilon_{auv}(\Omega_v +\Omega_{v}^B)+ \varepsilon_{cdb} V_{cl_b}B_a\partial_c G_{du} ) \mathcal{E}_u \nonumber\\
&&- \frac{e^2}{2} \int (f_2^0 + {f_2^0}^*)(\varepsilon_{auv}(\Omega_v +\Omega_{v}^B)+\varepsilon_{cdb} V_{cl_b}B_a\partial_c G_{du} ) \mathcal{E}_u.\nonumber\\
\end{eqnarray}
where in the above equations only those terms linear in $B$ should be retained. Consequently, dependent on the material, the magnetic and the electric field directions, we can simplify Eqs.~\ref{f1} to \ref{f20}, then by the substitution of Eqs.~\ref{f1} to \ref{f20}, in the above equations, we can derive different responses for different understudy systems.
\section{2D materials example}
In this section, 2D materials under a tilted magnetic field are the understudy system where for a 2D material, the solely nonzero Berry curvature component is $\Omega_z$ \cite{sodemann2015quantum}. For instance, $\Omega_z$, and even its dipole is nonzero for the massive Dirac points of TCIs \citep{sodemann2015quantum}, strained monolayer graphene, strained bilayer graphene \cite{battilomo2019berry}, and bilayer graphene under a steady in-plane magnetic field \cite{kheirabadi2022quantum}. 

We also assume that the electric field is in-plane; $E_\parallel=(E_x,E_y)$ and the applied magnetic field is $\mathbf{B}=(B_x,B_y,B_z)$. We have also ignored those terms include $\partial_{ijk} f_0^0$, $i,j,k=x, y$, because those are $\tau e \left|  \mathcal{E} \right| / \left| k \right|$ smaller compared to $\partial_{ij}f_0^0$ ones and those are in scale of $\tau^3$. Hence, we can simplify Eqs.~\ref{f1} to \ref{f20} for the assumed considered understudy system dependent on the indices $x$ and $y$. For example, we have $f_{1x}$ and $f_{1y}$ where $a$ is $x$ and $y$, respectively. The form of $f_{1x}$, $f_{1y}$, $f_{2x}$, $f_{2y}$, $f_{2x}^0$ and $f_{2y}^0$  is written in detail in Appendix.~\ref{A}. So, for $S=V_{cl_a}$, Eqs.~\ref{ja0} to \ref{ja1} are
\begin{eqnarray}\label{jx0}
j_x^0&&=-e \int (f_{2x}^0+f_{2y}^0) V_{cl_x} - \frac{e^2}{2} \int (f_{1x}+f_{1y}) (\Omega+\Omega^B) \mathcal{E}_y^*\nonumber\\
&& - \frac{e^2}{4} \int \big( f_0^0 + \frac{f_{2x}^0}{2}+\frac{f_{2y}^0}{2} \big) \big[ G_2 \big( \mathcal{E}_x \mathcal{E}_y^*+\mathcal{E}_x^* \mathcal{E}_y \big) + 2 G_1 \left| \mathcal{E}_y \right|^2 \big],\nonumber\\
\end{eqnarray}
\begin{eqnarray}
j_y^0&&=-e \int (f_{2y}^0+f_{2x}^0) V_{cl_y} + \frac{e^2}{2} \int (f_{1y}+f_{1x}) (\Omega+\Omega^B) \mathcal{E}_x^*\nonumber\\
&&+\frac{e^2}{4} \int \big( f_0^0 + \frac{f_{2y}^0}{2}+\frac{f_{2x}^0}{2} \big) \big[ G_1 \big( \mathcal{E}_x \mathcal{E}_y^*+\mathcal{E}_x^* \mathcal{E}_y \big) + 2 G_2 \left| \mathcal{E}_x \right|^2 \big],\nonumber\\
\end{eqnarray}
\begin{eqnarray}
j_x^2&&=-e \int (f_{2x}+f_{2y}) V_{cl_x} - \frac{e^2}{2} \int (f_{1x}+f_{1y}) (\Omega+\Omega^B) \mathcal{E}_y \nonumber\\
&&- \frac{e^2}{4} \int f_0^0  \big[ G_2 \big( \mathcal{E}_x \mathcal{E}_y+\mathcal{E}_x^* \mathcal{E}_y^* \big) + G_1 \big(\mathcal{E}_y ^2+{\mathcal{E}_y^*}^2 \big) \big],\nonumber\\
\end{eqnarray}
\begin{eqnarray}
j_y^2&&=-e \int (f_{2y}+f_{2x}) V_{cl_y} + \frac{e^2}{2} \int (f_{1y}+f_{1x}) (\Omega+\Omega^B) \mathcal{E}_x \nonumber\\
&&+ \frac{e^2}{4} \int f_0^0  \big[ G_1 \big( \mathcal{E}_x \mathcal{E}_y+\mathcal{E}_x^* \mathcal{E}_y^* \big) + G_2 \big(\mathcal{E}_x ^2+{\mathcal{E}_x^*}^2 \big) \big],\nonumber\\
\end{eqnarray}
\begin{eqnarray}\label{jx1}
j_x^1=&&-e \int (f_{1x}+f_{1y}) V_{cl_x} \nonumber\\
&&- e^2 \int \big( f_0^0 + \frac{1}{2}(f_{2x}^0 + {f_{2x}^0}^*+f_{2y}^0 + {f_{2y}^0}^*) \big)(\Omega+\Omega^B) \mathcal{E}_y,\nonumber\\
\end{eqnarray}
\begin{eqnarray}\label{jy1}
j_y^1=&&-e \int (f_{1y}+f_{1x}) V_{cl_y}\nonumber\\
&& + e^2 \int \big( f_0^0 + \frac{1}{2}(f_{2y}^0 + {f_{2y}^0}^*+f_{2x}^0 + {f_{2x}^0}^*) \big)(\Omega+\Omega^B) \mathcal{E}_x,\nonumber\\
\end{eqnarray}
where,
\begin{eqnarray}\label{G1}
G_1=\big(\partial_x G_{yy} - \partial_y G_{xy} \big),
\end{eqnarray}
\begin{eqnarray}\label{G2}
G_2=\big(\partial_x G_{yx} - \partial_y G_{xx} \big).
\end{eqnarray}
These equations are general solutions for 2D materials under any radiation type.
According to Eqs.~\ref{jx0} to \ref{jy1}, scattering time-independent currents are dependent on the quantum parameters. For instance, for the DC response and SHG one, $\int f_0^0 G_1$ and $\int f_0^0 G_2$ determine the intrinsic current. While, for the linear response, $\int f_0^0 (\Omega+\Omega^B)$ determines the intrinsic linear currents in 2D materials.   
In fact, for second-order terms, $\tau$ independent currents are the result of considering the Berry curvature correction, however for the linear current, it is a consequence of considering a nonzero $\Omega+\Omega^B$ factor.

For the ratchet or DC, we consider the general form of the current as the following form
\begin{eqnarray}
j_x^0&&=M_{1x} \mathcal{E}_x^* \mathcal{E}_y + M_{2x} \mathcal{E}_x \mathcal{E}_y^*+ M_{3x} \left| \mathcal{E}_x \right|^2 + M_{4x} \left| \mathcal{E}_y \right|^2,\nonumber\\
j_y^0&&=M_{1y} \mathcal{E}_x^* \mathcal{E}_y + M_{2y} \mathcal{E}_x \mathcal{E}_y^*+ M_{3y} \left| \mathcal{E}_x \right|^2 + M_{4y} \left| \mathcal{E}_y \right|^2,\nonumber\\
\end{eqnarray}
For the SHG response, there are 
\begin{eqnarray}
j_x^2=&&N_{1x} \mathcal{E}_x \mathcal{E}_y+N_{2x} \mathcal{E}_x^2+N_{3x} \mathcal{E}_y^2+N_{4x} \mathcal{E}_x^* \mathcal{E}_y^*+N_{5x} \mathcal{E}_x^{*^2}\nonumber\\
&&+N_{6x} \mathcal{E}_y^{*^2},\nonumber\\\
j_y^2=&&N_{1y} \mathcal{E}_x \mathcal{E}_y+N_{2y} \mathcal{E}_x^2+N_{3y} \mathcal{E}_y^2+N_{4y} \mathcal{E}_x^* \mathcal{E}_y^*+N_{5y} \mathcal{E}_x^{*^2}\nonumber\\
&&+N_{6y} \mathcal{E}_y^{*^2},\nonumber\\
\end{eqnarray}

For the linear response, the currents are written as 
\begin{eqnarray}
j_x^1=O_{1x} \mathcal{E}_x + O_{2x} \mathcal{E}_y,\nonumber\\
j_y^1=O_{1y} \mathcal{E}_x + O_{2y} \mathcal{E}_y,
\end{eqnarray}
where we leave expressions of the response tensors $M$, $N$, and $O$ in Appendix.~\ref{B}. In the Appendix.~\ref{C}, dependent on the point group of the material, the allowed (forbidden) conductivity tensor elements are also determined.

Under a tilted magnetic field, the in-plane magnetic field can also contribute in 2D materials Hamiltonian \cite{kheirabadi2016magnetic,kheirabadi2018cyclotron,  kheirabadi2021magnetic, kheirabadi2021current}, so for a 2D material under a tilted magnetic filed a $B_x$ and $B_y$ dependent BC, $\Omega (B_x, B_y)$, should be substituted in Eq.~\ref{M1x} to \ref{O2y} (Ref.~\cite{kheirabadi2022quantum}). Hence, a part of $\Omega$ which is dependent on the in-plane magnetic field comes from the Hamiltonian, while $\Omega^B$ comes from $\mathbf{A}_n^{\prime}(B)$ in Eq.~\ref{Eq1}. 

\textbf{Candidate material: The massive Dirac points of TCIs}

In this section, we study the effect of an out-of-plane magnetic field on the deduced currents in a 2D system, TCI. The low energy Hamiltonian for the massive Dirac points of TCIs is \cite{sodemann2015quantum}
\begin{eqnarray}\label{Hamil}
H= v_x k_x \sigma_y - \xi v_y k_y \sigma_x  + \xi \alpha k_y + \beta \sigma_z,
\end{eqnarray}
where $\sigma_i$ determines Pauli matrices, $\xi$ is the valley index, $\beta$ is the size of the gap, and $\alpha$ determines the tilt in the Dirac cones \cite{sodemann2015quantum}. This Hamiltonian is also valid for the strained monolayer transition-metal dichalcogenides \citep{sodemann2015quantum}. We can show that the dispersion relation, the Berry curvature, $\Omega$, and the BCP tensor, Eq. \ref{Gtensor}, for this Hamiltonian are 
\begin{eqnarray}
E (k_x, k_y, \xi) = \alpha \xi k_y + s \zeta, 
\end{eqnarray}
\begin{eqnarray}
\Omega = \frac{s}{2}\frac{\xi v_x v_y \beta}{\zeta^3},
\end{eqnarray}
and,
\begin{eqnarray}
G = \frac{s}{4 \zeta^5} \begin{pmatrix}
v_x^2 (v_y^2 k_y^2  + \beta^2) & -v_x^2 v_y^2 (k_x k_y) \\
-v_x^2 v_y^2 (k_x k_y) & -v_y^2 ( v_x^2 k_x^2 + \beta^2),
\end{pmatrix}
\end{eqnarray}
where $s=-1(+1)$ for the valence (conduction) band and $\zeta=(v_x^2 k_x^2  + v_y^2 k_y^2  + \beta^2)^{1/2}$.  
Consequently, we can show that $G_1$
and $G_2$, Eqs. \ref{G1} and \ref{G2}, are
\begin{eqnarray}
G_1&&=\frac{-s}{2\zeta^5} v_x^2 v_y^2 k_x,\nonumber\\
G_2&&=\frac{s}{2\zeta^5} v_x^2 v_y^2 k_y.
\end{eqnarray}
Also, according to Eqs.~\ref{Eq1}, \ref{FtensorS}, and \ref{FtensorO}, we can show that $\Omega^B$ for the conduction and valence bands are equal to
\begin{eqnarray}
\Omega^B= \frac{v_x^2 v_y^2 \beta^2}{\zeta^4} B_z.
\end{eqnarray}
We note that the AOP dominates over the ASP when the Fermi energy approaches the degenerate band points or band-crossing points. Subsequently, we overlook the ASP contribution below. In addition, making use of the periodic boundary condition, we can show that for a general function $g(k_x,k_y)$
\begin{eqnarray}\label{int1}
\int g(k_x,k_y) \partial_{a}f_0^0 = -\int \partial_{a}g(k_x,k_y) f_0^0,
\end{eqnarray}
\begin{eqnarray}\label{int2}
\int g(k_x,k_y) \partial_{ab}f_0^0 = \int \partial_{ab}g(k_x,k_y) f_0^0.
\end{eqnarray}
For example, according to Eq.~\ref{int1}, $\int \Omega \partial_i f_0^0 = - \int  f_0^0 \partial_i \Omega$ where at zero temperature $f_0^0 = \Theta[\mu-E(k_x, k_y, \xi)]$ ($\mu$ is the chemical potential). 

For nonlinear responses in TCIs, $M$ and $N$ related responses, we can show that there are two types of nonzero responses. The first group which has been studied before under the name of the effect of Berry curvature dipole (BCD) is proportional to $\int f_0^0 \partial_i \Omega$. In TCIs, only BCD in the $y$ direction survives. Because $\partial_i \Omega  \propto s \xi k_i / \zeta^{5}$. Consequently, the result of this integral is zero for $i=x$ because if $k_x \to -k_x$, then $E(-k_x, k_y, \xi)= E(k_x, k_y, \xi)$, and $s \xi k_x / \zeta^{5} \to -s  \xi k_x / \zeta^{5}$, so $\int_{0}^{+\infty} \Theta[\mu-E(k_x, k_y, \xi)] \partial_x \Omega dk_x=-\int_{- \infty}^{0} \Theta[\mu-E(k_x, k_y, \xi)] \partial_x \Omega dk_x$. However, for $i=y$, if $k_y \to - k_y$, $E(k_x, -k_y, \xi) = E(k_x, k_y, -\xi)$, so $s \xi k_y / \zeta^{5} \to s  \xi k_y / \zeta^{5}$ and consequently, BCD in the $y$ direction, $\int  f_0^0 \partial_y \Omega$,  survives and it originates a second order (nonlinear) $\tau$ dependent quantum Hall current \cite{sodemann2015quantum}. This nonzero BCD results in a quantum Hall effect for materials that are invariant under time-reversal symmetry \cite{sodemann2015quantum, battilomo2019berry} and even when the time-reversal symmetry is broken by in-plane magnetic field \citep{kheirabadi2022quantum}.
With the same analysis, also according to the time-reversal symmetry analysis and the dependency of integrand to $k_x$ and $k_y$, we can show that considering $\int V_{cl_i} \Omega \partial_{jk} f_0^0 = \int \Theta[\mu-E(k_x, k_y, \xi)] \partial_{jk}(V_{cl_i} \Omega)$  results in the second group of surviving nonlinear currents in TCIs. These types of responses are in order of $\tau^2$ and those are proportional to the $B_z$. These nonzero integrals are $\int f_0^0 \partial_{xy}(V_{cl_x} \Omega)$, $\int f_0^0 \partial_{xx}(V_{cl_y} \Omega)$, and $\int f_0^0 \partial_{yy}(V_{cl_y} \Omega)$. Note that for the planar Hall effect (when the applied magnetic field is in-plane), a nonzero Berry curvature dipole, $\int f_0^0 \partial_i \Omega(B_x, B_y)$ determines a nonzero second order Hall effect \cite{kheirabadi2022quantum}. While, for a perpendicular magnetic field, $ B_z \int f_0^0 \partial_{ij}(V_{{cl}_{k}} \Omega_z)$ is the determining parameter to have the additional nonlinear Hall current, means that the deduced current is dependent on the multiplication of the classical velocity and BC which is a quantity in the quantum mechanic. Finally, it is worthy to mention that for $\alpha=0$, two valleys cancel out the nonlinear responses of each other. 

For the linear responses that survive even at $\alpha=0$, the major effect belongs to the classical responses originating from $\int V_{cl_i} \partial_i f_0^0$. We can show that these integrals are proportional to the integrals of an even function of $k_i$, so those survive after integration. Furthermore, nonzero classical Hall responses which have a $B_z$ coefficient are proportional to $\int V_{cl_i} V_{cl_j} \partial_{ij} f_0^0$.
$\int f_0^0 \Omega^B$ that is related to the intrinsic Hall conductivity also survives for TCIs.

\begin{widetext} 
\begin{figure*}
   \centering   
   \includegraphics[scale=0.47]{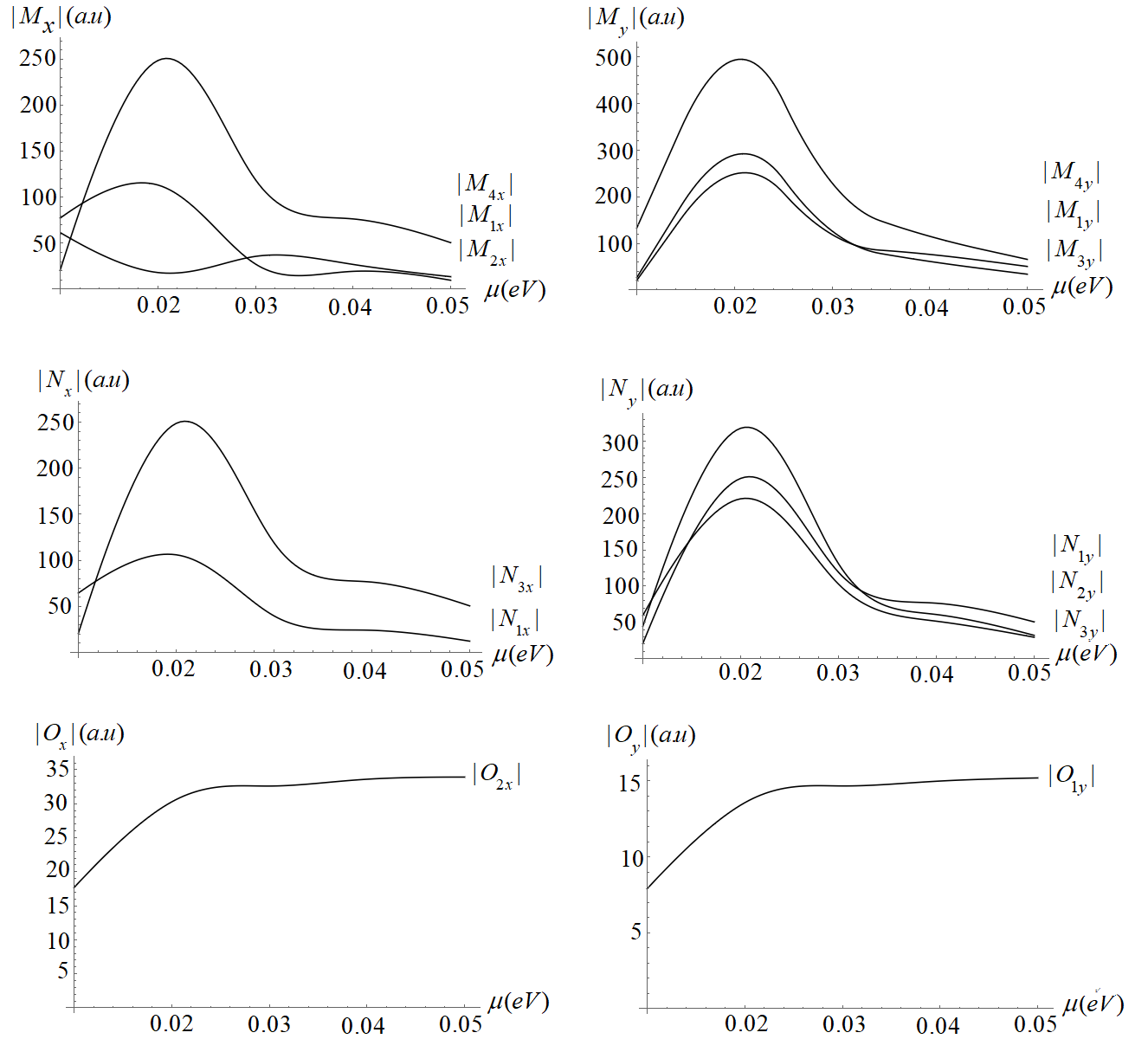}
   \caption{The absolute values of $M$, $N$ and $O$ prefactors in the $x$ and $y$ directions as a function of the chemical potential $\mu$ for SnTe under a $10$ T perpendicular magnetic field and by considering $\omega \tau=1$.} 
    \label{diag}
\end{figure*}
\end{widetext}

The integrals related to $M$ (Eqs.~\ref{M1x} to \ref{M4y}), $N$ (Eqs.~\ref{N1x} to \ref{N6y}), and $O$ (Eqs.~\ref{O1x} to \ref{O2y}) prefactors could also be solved for two bands of the massive Dirac points. 
For instance, for SnTe, in the Hamiltonian of Eq.~\ref{Hamil}, we have $v_x \approx v_y \approx 4 \times 10^5$ m/s, $\beta \approx 10$ meV, $\alpha \approx 0.1 v_x$, and $\tau=0.2$ ps \citep{sodemann2015quantum,jiang2012landau}. Assuming a perpendicular magnetic field $B_z=10$ T, and $\omega \tau =1$ results in the following diagrams for $M$, $N$, and $O$ prefactors for the massive Dirac points of SnTe dependent on the chemical potential, $\mu$ (Fig.~\ref{diag}).
Additionally,  in keeping with numerical calculations, for the linear responses, the major effects belong to $O_{1x}$ and $O_{2y}$ which are in order of nA/V. In a chemical potential of $10$ meV, for the mentioned parameters of SnTe, for an electric field $E(t)$ with amplitude $7$ kVcm$^{-1}$, we can show that the deduced quantum current is in order of $10$ $\mu$A/cm. The magnitude of the Hall responses of the current resulted by both velocity and Berry curvature terms / quantum term, $\tau e B_z\int f_0^0 \partial_{ij} (V_{cl_k} \Omega)/ \int f_0^0 \partial_y \Omega$, for ($i \neq j$, $k=x$), ($i=j= y$, $k=y$), ($i=j=x$, $k=y$) are also 1.5, 3.4, and 0.4,  respectively. These quantities that demonstrate the magnitude of the current resulted by both velocity and BC currents in contrast to the quantum current deduced by BCD in SnTe may be improved linearly by the change of the perpendicular magnetic field. The intrinsic linear quantum current which is proportional to $\int \Omega^B f_0^0$ is likewise so as 0.1 $\mu$A/cm and two orders of magnitude smaller than the currents that originate from the BCD and the current resulted by both velocity and BC.  

Moreover, for $B_z=0$, the outcomes of this work are in agreement with the outcomes expected by Ref.~\cite {sodemann2015quantum}. For example, assuming $B_z=0$ in Eqs.~\ref{M1x} to \ref{N6y} outcomes in nonzero $M_{4x}$, $M_{1y}$, $N_{3x}$ and $N_{1y}$ prefactors. These prefactors determine $\chi_{xyy}$ and $\chi_{yyx}$ for the massive Dirac points of TCIs which are nonzero according to Ref.~\citep{sodemann2015quantum}.  

In addition, it is notable that all of the nonlinear responses of TCIs have the quantum origin $\Omega$. The linear response, the first terms of Eqs.~\ref{O1x} and \ref{O2y}, is the major effect. Hence, by choosing the electric field solely in the $x$ or only in the $y$ direction, the Hall current will have the quantum characteristics. This happens because the magnitude of the only linear classical Hall response, the first terms of Eqs.~ \ref{O2x} and \ref{O1y}, are comparable with nonlinear quantum Hall responses (Fig.~\ref{diag}).
We also point out that because the massless Dirac points of TCIs have a zero Berry curvature \cite{sodemann2015quantum}, considering two other massive Dirac points changes the classical currents only and Hall currents that are BC dependent are similar to what have been predicted for the massive Dirac points in TCIs.

\section{Conclusion}
In previous studies, it was found that non-linear Hall currents have a dependence on the BCD and the effect of an in-plane magnetic field on the BCD has also been identified \cite{sodemann2015quantum, kheirabadi2022quantum}. Here we extend previous results, and we derive the general form of the current up to the second order in the electrical field for materials under THz or microwave radiation and a tilted magnetic field. We have considered the effect of a tilted steady magnetic field on electrons of 2D and 3D materials by the use of the Boltzmann kinetic equation, and we demonstrate the form of the intrinsic responses in this regime. It has also been modelled how the tilted magnetic field affects both intrinsic and extrinsic currents. The conductivity tensor for 2D materials is also derived due to the importance of application. Within a candidate material, TCI, we found two types of surviving terms in SHG and ratchet responses; the first is proportional to the BCD, and it has a quantum origin and has been studied before in Ref.~\cite{sodemann2015quantum}. The second term is a current resulted by both velocity and BC, and it is proportional to the perpendicular component of the tilted magnetic field, and we show that these terms vanish for $\alpha=0$. A quantum intrinsic Hall current based on $B_z$ is predicted for the linear response, along with classical currents. A numerical demonstration is also used to demonstrate the magnitude of the conductivity tensor elements, and the intrinsic magnetic field-dependent Hall current for SnTe as a TCI example. 
\section{Acknowledgement}
N.~K thanks E.~McCann for useful comments.
\appendix
\section{The form of $f$ functions for $a=x$ and $a=y$}\label{A}
If we consider that $a=x$ in Eqs.~\ref{f1} to \ref{f20}, we can write 
\begin{widetext}
\begin{eqnarray}\label{f1x}
f_{1x}=\frac{\tau e}{1+i \omega \tau} \bigg[&& \bigg( \big(1+ eB_z \Omega_z \big) \partial_x f_0^0 - \tau e B_z (V_{cl_y}  \partial_{xx} f_0^0-V_{cl_x}  \partial_{xy} f_0^0 )\bigg) \mathcal{E}_x
 + \frac{\tau e }{1+i \omega \tau} V_{cl_y} B_z (\partial_{xx} f_0^0 \mathcal{E}_x+\partial_{xy} f_0^0 \mathcal{E}_y) \bigg],\nonumber\\
\end{eqnarray}
\begin{eqnarray}\label{f2x}
f_{2x}=\frac{\tau^2 e^2}{2(1+i \omega \tau)(1+2i \omega \tau)} \bigg[ &&\big(1 -2 e B_z \Omega_z \big) (\partial_{xy} f_0^0  \mathcal{E}_x \mathcal{E}_y + \partial_{xx} f_0^0  \mathcal{E}_x^2)
 + \frac{\tau e^2 }{2(1 + 2 i \omega \tau)} B_z
\big[ (\partial_x-\partial_y) (G_{xx} \mathcal{E}_x^2 + G_{xy} \mathcal{E}_x \mathcal{E}_y) \big] \partial_x f_0^0 \bigg],\nonumber\\
\end{eqnarray}
\begin{eqnarray}\label{f20x}
f_{2x}^0=\frac{\tau^2 e^2}{2(1+i \omega \tau)}  \bigg( \big( 1 - 2 e B_z \Omega_z \big) \mathcal{E}_x^* \mathcal{E}_y \partial_{xy}f_0^0 + \big(1 - 2 e B_z \Omega_z \big) \left| \mathcal{E}_x \right|^2 \partial_{xx}f_0^0 \bigg)\nonumber\\
+ \bigg[ \tau e V_{cl_y} B_z - \frac{\tau e^2}{4} B_z\big[ G_1 (\mathcal{E}_x^* \mathcal{E}_y + \mathcal{E}_x \mathcal{E}_y^*) + 2 G_2 \left| \mathcal{E}_x \right|^2 \big] \bigg]\partial_x f_0^0,
\end{eqnarray}
Also, considering $a=y$ in Eqs.~\ref{f1} to \ref{f20}, we have  
\begin{eqnarray}\label{f1y}
f_{1y}=\frac{\tau e}{1+i \omega \tau} \bigg[&& \bigg( \big(1 - eB_z \Omega_z \big) \partial_y f_0^0 + \tau e B_z ( V_{cl_y} \partial_{yx} f_0^0 -V_{cl_x} \partial_{yy} f_0^0)\bigg) \mathcal{E}_y 
- \frac{\tau e }{1+i \omega \tau} B_z V_{cl_x} (\partial_{yy} f_0^0 \mathcal{E}_y + \partial_{yx} f_0^0 \mathcal{E}_x) \bigg],
\end{eqnarray}
\begin{eqnarray}\label{f2y}
f_{2y}=\frac{\tau^2 e^2}{2(1+i \omega \tau)(1+2i \omega \tau)} \bigg[ &&\big(1 - 2 e B_z \Omega_z \big) (\partial_{yx} f_0^0  \mathcal{E}_x \mathcal{E}_y + \partial_{yy} f_0^0 \mathcal{E}_y^2)
- \frac{\tau e^2 }{2(1+ 2 i \omega \tau)} B_z  \big[ (\partial_x-\partial_y) (G_{yy} \mathcal{E}_y^2 + G_{yx} \mathcal{E}_x \mathcal{E}_y) \big] \partial_y f_0^0 \bigg],\nonumber\\
\end{eqnarray}
\begin{eqnarray}\label{f20y}
f_{2y}^0&&=\frac{\tau^2 e^2}{2(1+i \omega \tau)} \bigg( \big( 1 - 2 e B_z \Omega_z \big) \mathcal{E}_x\mathcal{E}_y^*  \partial_{yx}f_0^0 + \big( 1- 2 e B_z \Omega_z \big) \left| \mathcal{E}_y \right|^2 \partial_{yy}f_0^0\bigg)\nonumber\\
&&+ \bigg[ \tau e V_{cl_x} B_z -\frac{\tau e^2}{4} B_z\big[ G_2 (\mathcal{E}_x^* \mathcal{E}_y+ \mathcal{E}_x \mathcal{E}_y^*) + 2 G_1 \left| \mathcal{E}_y \right|^2 \big] \bigg]\partial_y f_0^0.
\end{eqnarray}
\end{widetext}
\section{Expressions of the response tensors for ratchet or DC current}\label{B}
\begin{widetext}
The expressions of the response tensors for the ratchet or DC current are found as 
\begin{eqnarray}\label{M1x}
M_{1x}=- \frac{e^2}{2} \int &&\frac{\tau^2 e}{(1+i \omega \tau)}  V_{cl_x}  \big( 1 - 2 e B_z \Omega \big) \partial_{xy}f_0^0
 - \frac{\tau e}{2} B_z V_{cl_x} (G_1 \partial_{x}f_0^0+G_2 \partial_{y}f_0^0) 
+ \frac{1}{2}\big( f_0^0 + \frac{1}{2} \tau e B_z (V_{cl_y}\partial_x f_0^0 + V_{cl_x}\partial_y f_0^0) \big) G_2 ,\nonumber\\
\end{eqnarray}
\begin{eqnarray}\label{M2x}
M_{2x}=\frac{e^2}{2} \int && - \frac{\tau^2 e}{1+i \omega \tau}V_{{cl}_x}(1- 2 e B_z \Omega) \partial_{yx} f_0^0
+\frac{\tau e}{2} B_z V_{cl_x}( G_1 \partial_{x}f_0^0+ G_2 \partial_{y}f_0^0) \nonumber\\
&& - \frac{\tau e}{1+i \omega \tau} \bigg( \big(\Omega+\Omega^B+ eB_z \Omega^2 \big) \partial_x f_0^0 +\tau e B_z \Omega \big( V_{cl_y}  \partial_{xx} f_0^0 - V_{cl_x}  \partial_{xy} f_0^0 \big)-\frac{\tau e}{1+i \omega \tau} B_z \Omega \big( V_{cl_x}   \partial_{yx} f_0^0 - V_{cl_y}   \partial_{xx} f_0^0 \big) \bigg)\nonumber\\
&&- \frac{1}{2}\big( f_0^0 + \frac{1}{2} \tau e  B_z (V_{cl_y} \partial_x f_0^0 + V_{cl_x} \partial_y f_0^0) \big) G_2,\nonumber\\
\end{eqnarray}
\begin{eqnarray}\label{M3x}
M_{3x} = \frac{\tau e^3}{2}  \int  B_z V_{cl_x}\  G_2 \partial_x f_0^0 - \frac{\tau}{1 + i \omega \tau } V_{cl_x} \big( 1 - 2 e B_z \Omega_z \big)  \partial_{xx} f_0^0,  
\end{eqnarray}
\begin{eqnarray}\label{M4x}
M_{4x}=-\frac{e^2}{2} \int && \frac{\tau^2 e}{1+i \omega \tau} V_{cl_x} \big(1 - 2 e B_z \Omega_z \big) \partial_{yy} f_0^0+ \frac{\tau^2 e^2 }{(1+i \omega \tau)^2}  B_z \Omega \big( V_{cl_y} \partial_{xy} f_0^0 - V_{cl_x} \partial_{yy} f_0^0 \big)
\nonumber\\
&&+ \big( f_0^0 + \frac{1}{2}\tau e B_z (V_{cl_y} \partial_x f_0^0 - V_{cl_x} \partial_y f_0^0 )\big) G_1\nonumber\\
&&+\frac{\tau e }{1+i \omega \tau}\big[(\Omega+\Omega^B-e B_z \Omega^2) \partial_y f_0^0 +\tau e  B_z \Omega \big(V_{cl_y}  \partial_{yx}f_0^0 -V_{cl_x}  \partial_{yy}f_0^0 \big) \big],
\end{eqnarray}
\begin{eqnarray}\label{M1y}
M_{1y}=  \frac{e^2}{2} \int && - \frac{\tau^2 e}{1+i \omega \tau} V_{cl_y}(1- 2 e B_z \Omega ) \partial_{xy} f_0^0+ \frac{\tau e}{2} B_z V_{cl_y}  ( G_2 \partial_y f_0^0+G_1 \partial_{x} f_0^0 )\nonumber\\ 
&&+ \frac{\tau e}{1+i \omega \tau} \bigg( \big(\Omega+ \Omega^B - e B_z \Omega^2 \big) \partial_y f_0^0 + \tau e  B_z \Omega \big(V_{cl_y}  \partial_{yx} f_0^0 - V_{cl_x}  \partial_{yy} f_0^0\big) + \frac{\tau e}{1+i \omega \tau} B_z \Omega \big(V_{cl_y}  \partial_{xy} f_0^0 - V_{cl_x}  \partial_{yy} f_0^0 \big) \bigg) \nonumber\\
&& + \frac{1}{2} \big( f_0^0 + \frac{1}{2} \tau e B_z (V_{cl_x}  \partial_y f_0^0 + V_{cl_y}  \partial_x f_0^0 \big) \big) G_1,
\end{eqnarray}

\begin{eqnarray}\label{M2y}
M_{2y}=- \frac{e^2}{2} \int \frac{\tau^2 e }{(1+i \omega \tau)} V_{cl_y} \big( 1 - 2 e B_z \Omega \big) \partial_{yx}f_0^0 - \frac{\tau e}{2} B_z V_{cl_y} ( G_2 \partial_y f_0^0+G_1 \partial_x f_0^0) - \frac{1}{2} \big( f_0^0 + \frac{1}{2} \tau e  B_z (V_{cl_x} \partial_y f_0^0 + V_{cl_y} \partial_x f_0^0)\big) G_1,\nonumber\\
\end{eqnarray}

\begin{eqnarray}\label{M3y}
M_{3y} =  - \frac{e^2}{2} \int && \frac{\tau^2 e}{1+i \omega \tau} V_{cl_y} \big( 1 - 2 e B_z \Omega_z \big) \partial_{xx} f_0^0 + \frac{\tau^2 e^2}{(1+i \omega \tau)^2} B_z  \Omega \big( V_{cl_x}  \partial_{yx} f_0^0 - V_{cl_y}  \partial_{xx} f_0^0 \big) 
\nonumber\\
&& -\big( f_0^0 + \frac{1}{2} \tau e B_z (V_{cl_x} \partial_y f_0^0 - V_{cl_y} \partial_x f_0^0)\big) G_2 \nonumber\\
&& -\frac{\tau e}{1+i \omega \tau} \big[ \big( \Omega + \Omega^B + e B_z \Omega^2 \big) \partial_x f_0^0 - \tau e  B_z \Omega \big( V_{cl_y}  \partial_{xx} f_0^0  - V_{cl_x}  \partial_{xy} f_0^0 \big)\big],  
\end{eqnarray}
\begin{eqnarray}\label{M4y}
M_{4y}= \frac{\tau e^3}{2}  \int  B_z V_{cl_y} G_1 \partial_y f_0^0 - \frac{\tau}{1 + i \omega \tau} V_{cl_y} (1-2 e B_z \Omega_z) \partial_{yy} f_0^0.
\end{eqnarray}

For the response tensors for SHG current, there are
\begin{eqnarray}\label{N1x}
N_{1x}=\frac{- e^2}{2}\int && \frac{ \tau^2 e}{(1+i \omega \tau)(1+2i \omega \tau)} V_{cl_x} \bigg[ \big(1 -2 e B_z \Omega \big) \big(\partial_{xy} f_0^0 +\partial_{yx} f_0^0 \big) + \frac{\tau e^2 }{2 (1 + 2 i \omega \tau)} B_z \big[ (\partial_x-\partial_y)  G_{xy} \partial_x f_0^0  - (\partial_x-\partial_y)  G_{yx} \partial_y f_0^0 \big] \bigg]\nonumber\\
&& + \frac{\tau e}{1+i \omega \tau} \bigg[ \bigg( \big(\Omega+\Omega^B+ eB_z \Omega^2 \big) \partial_x f_0^0 + \tau e  B_z \Omega \big( V_{cl_x}  \partial_{xy} f_0^0 - V_{cl_y}  \partial_{xx} f_0^0\big)\bigg) \bigg]\nonumber\\
&& +\frac{1}{2} f_0^0 G_2- \frac{\tau^2 e^2}{(1+i \omega \tau)^2} B_z \Omega \big( V_{cl_x}  \partial_{yx} f_0^0 -  V_{cl_y}  \partial_{xx} f_0^0 \big),
\end{eqnarray}
\begin{eqnarray}\label{N2x}
N_{2x}=\frac{-\tau^2 e^3}{2 (1+i \omega \tau)(1+2i \omega \tau)}  \int \frac{\tau e^2 }{2(1+2i\omega \tau)} B_z V_{cl_x}  (\partial_x-\partial_y) G_{xx} \partial_x f_0^0 + V_{cl_x} \big( 1 - 2 e B_z \Omega_z\big) \partial_{xx} f_0^0,
\end{eqnarray}
\begin{eqnarray}\label{N3x}
N_{3x}= -\frac{e^2}{2}\int && \frac{\tau^2 e}{(1+i \omega \tau)(1+2i \omega \tau)} V_{cl_x} (1-2 e B_z \Omega_z ) \partial_{yy}f_0^0 + \frac{\tau^2 e^2}{(1+i \omega \tau)^2} B_z \Omega \big( V_{cl_y} \partial_{xy} f_0^0 -V_{cl_x} \partial_{yy} f_0^0 \big)\nonumber\\
&& + \frac{1}{2} f_0^0 G_1- \frac{\tau^3 e^3 }{2(1+i \omega \tau)(1+2i \omega \tau)^2} B_z V_{cl_x} (\partial_x-\partial_y) G_{yy} \partial_y f_0^0\nonumber\\
&& + \frac{\tau e }{(1+i \omega \tau)} \big[ \big(\Omega + \Omega^B- e B_z\Omega^2 \big) \partial_y f_0^0+ \tau e B_z \Omega \big(V_{cl_y}  \partial_{yx}f_0^0 - V_{cl_x}  \partial_{yy}f_0^0\big) \big],
\end{eqnarray}
\begin{eqnarray}\label{N4x}
N_{4x}=-\frac{e^2}{4}\int f_0^0 G_2,
\end{eqnarray}
\begin{eqnarray}\label{N5x}
N_{5x}=0,
\end{eqnarray}
\begin{eqnarray}\label{N6x}
N_{6x}=-\frac{e^2}{4}\int f_0^0 G_1,
\end{eqnarray}
\begin{eqnarray}\label{N1y}
N_{1y}=\frac{e^2}{2}\int && \frac{- \tau^2 e}{(1+i \omega \tau)(1+2i \omega \tau)} V_{cl_y} \bigg[ \big(1 - 2 e B_z \Omega \big) \big(\partial_{yx} f_0^0 +\partial_{xy} f_0^0 \big) - \frac{\tau e^2 }{2(1+ 2 i \omega \tau)} B_z [(\partial_x-\partial_y)  G_{yx} \partial_y f_0^0 -(\partial_x-\partial_y)  G_{xy} \partial_x f_0^0 ]\bigg]\nonumber\\
&& + \frac{\tau e}{(1+i \omega \tau)} \bigg[ \bigg( \big(\Omega+\Omega^B - eB_z \Omega^2 \big) \partial_y f_0^0 + \tau e B_z \Omega \big( V_{cl_y} \partial_{yx} f_0^0 - V_{cl_x} \partial_{yy} f_0^0 \big) \bigg)  \bigg]\nonumber\\
&&+ \frac{1}{2} f_0^0 G_1 + \frac{\tau^2 e^2}{(1+i \omega \tau)^2 }  B_z \Omega \big( V_{cl_y}  \partial_{xy}f_0^0 - V_{cl_x}  \partial_{yy}f_0^0  \big),
\end{eqnarray}
\begin{eqnarray}\label{N2y}
N_{2y}= -\frac{e^2}{2}\int && \frac{\tau^2 e}{(1+i \omega \tau)(1+2i \omega \tau)} V_{cl_y} (1-2 e B_z \Omega_z) \partial_{xx} f_0^0 + \frac{\tau^2 e^2 }{(1+i \omega \tau)^2}B_z \Omega  \big( V_{cl_x}  \partial_{yx} f_0^0 - V_{cl_y}  \partial_{xx} f_0^0\big)\nonumber\\
&& -\frac{1}{2} f_0^0 G_2 + \frac{\tau^3 e^3}{ 2 (1+i \omega \tau)(1+2i \omega \tau)^2} B_z V_{cl_y} (\partial_x-\partial_y) G_{xx} \partial_x f_0^0\nonumber\\
&&-\frac{\tau e }{1+ i \omega \tau}\big[(\Omega+\Omega^B+e B_z\Omega^2) \partial_x f_0^0 -\tau e  B_z \Omega \big(V_{cl_x}  \partial_{xy} f_0^0 - V_{cl_y}  \partial_{xx} f_0^0\big)\big],
\end{eqnarray}
\begin{eqnarray}\label{N3y}
N_{3y}=\frac{\tau^2 e^3}{2(1+i \omega \tau)(1+2i \omega \tau)}   \int \frac{\tau e^2}{2(1+2i \omega \tau)} V_{cl_y} B_z (\partial_x-\partial_y) G_{yy}  \partial_y f_0^0 - V_{cl_y} \big(1- 2 e B_z \Omega_z \big) \partial_{yy} f_0^0,
\end{eqnarray}
\begin{eqnarray}\label{N4y}
N_{4y}= \frac{e^2}{4} \int f_0^0 G_1,
\end{eqnarray}
\begin{eqnarray}\label{N5y}
N_{5y}=\frac{e^2}{4} \int f_0^0 G_2,
\end{eqnarray}
\begin{eqnarray}\label{N6y}
N_{6y}=0.
\end{eqnarray}
The expressions for the linear response tensors are given as
\begin{eqnarray}\label{O1x}
O_{1x}= - \frac{\tau e^2}{1+i \omega \tau} \int V_{cl_x} \bigg( \big(1+ eB_z \Omega \big) \partial_x f_0^0 - \tau e  B_z (V_{cl_y}\partial_{xx} f_0^0 - V_{cl_x}\partial_{xy} f_0^0 )-\frac{\tau e }{1+i \omega \tau}  B_z \big( V_{cl_x} \partial_{yx}f_0^0 - V_{cl_y} \partial_{xx}f_0^0 \big)\bigg),\nonumber\\
\end{eqnarray}
\begin{eqnarray}\label{O2x}
O_{2x}=- e^2 \int && \frac{\tau^2 e}{(1+i \omega \tau)^2} V_{cl_x} B_z \big( V_{cl_y}  \partial_{xy} f_0^0 - V_{cl_x}  \partial_{yy} f_0^0 \big)
+ \big( (\Omega+\Omega^B) f_0^0 + \tau e B_z \Omega (V_{cl_y} \partial_x f_0^0 +  V_{cl_x} \partial_y f_0^0) \big)\nonumber\\
&&+\frac{\tau }{1+i \omega \tau} V_{cl_x} \big[(1-eB_z \Omega) \partial_y f_0^0 + \tau e B_z \big(V_{cl_y}  \partial_{yx} f_0^0 - V_{cl_x}  \partial_{yy} f_0^0  \big)\big],
\end{eqnarray}
\begin{eqnarray}\label{O1y}
O_{1y}= -e^2 \int && \frac{\tau^2 e}{(1+i \omega \tau)^2} V_{cl_y} B_z \big( V_{cl_y}  \partial_{xx}f_0^0 - V_{cl_x}  \partial_{yx}f_0^0 \big)-\big( (\Omega+\Omega^B ) f_0^0 + \tau e B_z \Omega (V_{cl_y} \partial_x f_0^0 + V_{cl_x} \partial_y f_0^0) \big)\nonumber\\
&&+\frac{\tau  }{1+i \omega \tau} V_{cl_y} \big[ (1+e B_z \Omega)\partial_x f_0^0 - \tau e  B_z \big(V_{cl_y} \partial_{xx} f_0^0 - V_{cl_x} \partial_{xy}f_0^0 \big) \big], 
\end{eqnarray}
\begin{eqnarray}\label{O2y}
O_{2y}= - \frac{\tau e^2}{1+i \omega \tau}\int  V_{cl_y} \bigg( \big(1 - eB_z \Omega \big) \partial_y f_0^0 + \tau e  B_z \big(V_{cl_y}\partial_{yx} f_0^0 - V_{cl_x}\partial_{xx} f_0^0\big)+ \frac{\tau e}{1+i \omega \tau}B_z \big( V_{cl_y} \partial_{xy} f_0^0 - V_{cl_x} \partial_{yy} f_0^0 \big) \bigg ).\nonumber\\
\end{eqnarray}
\end{widetext}
\section{Constraints on $M$ and $N$ tensors by the generators of the point groups}\label{C}
\begin{table*}[htbp]
	\renewcommand\arraystretch{2}
	\caption{List of constraints on $M$ and $N$ tensors by the generators of point groups. The allowed (forbidden) conductivity tensor elements are indicated by $\checkmark$ ($\times$). }\label{mgp2}
	\begin{tabular*}{17cm}{@{\extracolsep{\fill}}p{0.80cm}cccccccccc}
		\hline\hline
         & $\mathcal{P}$ & $C_2^y$ & $C_2^z$ & $\mathcal{P} C_2^y$ & $\mathcal{P} C_2^z$ & $C_3^z$ & $C_4^z$ & $\mathcal{P} C_4^z$ & $C_4^{z}\sigma_v$ 
          \\
		\hline
        $M_{1x}$
        &$\times$ & $\checkmark$ &$\times$ &$\times$  &  $\checkmark$    & $\checkmark$ & $\times$ & $\checkmark$& $\checkmark$ 
        \\
        \hline
	  $M_{2x}$
	  & $\times$ &$\checkmark$  &$\times$ &$\times$ &  $\checkmark$  & $\checkmark$ & $\times$ & $\checkmark$ &$\checkmark$
	  \\
	  \hline
	  $M_{3x}$ & $\times$ & $\times$ & $\times$ & $\checkmark$ &  $\checkmark$  & $\checkmark$ & $\times$ & $\times$ &$\checkmark$ 
	  \\
	  \hline
	  $M_{4x}$ & $\times$ & $\times$ & $\times$ & $\checkmark$ &  $\checkmark$  & $\checkmark$ & $\times$ & $\times$ &$\checkmark$
 \\ \hline
 $M_{1y}$
        &$\times$ & $\times$ &$\times$ &$\checkmark$  &  $\checkmark$    & $\checkmark$ & $\checkmark$ & $\times$& $\checkmark$
        \\
        \hline
	  $M_{2y}$
	  & $\times$ &$\times$  &$\times$ &$\checkmark$ &  $\checkmark$  & $\checkmark$ & $\checkmark$ & $\times$ &$\checkmark$
	  \\
	  \hline
	  $M_{3y}$ & $\times$ & $\checkmark$ & $\times$ & $\times$ &  $\checkmark$  & $\checkmark$ & $\times$ & $\times$ &$\checkmark$
	  \\
	  \hline
	  $M_{4y}$ & $\times$ & $\checkmark$ & $\times$ & $\times$ &  $\checkmark$  & $\checkmark$ & $\times$ & $\times$ &$\checkmark$\\
	  \hline
        $N_{1x}$
        &$\times$ & $\checkmark$ &$\times$ &$\times$  &  $\checkmark$    & $\checkmark$ & $\times$ & $\checkmark$& $\checkmark$
        \\
        \hline
	  $N_{2x}$
	  & $\times$ & $\times$ & $\times$ & $\checkmark$ &  $\checkmark$  & $\checkmark$ & $\times$ & $\times$ &$\checkmark$
	  \\
	  \hline
	  $N_{3x}$ & $\times$ & $\times$ & $\times$ & $\checkmark$ &  $\checkmark$  & $\checkmark$ & $\times$ & $\times$ &$\checkmark$
	  \\
	  \hline
	  $N_{4x}$ & $\times$ & $\checkmark$ &$\times$ &$\times$  &  $\checkmark$    & $\checkmark$ & $\times$ & $\checkmark$& $\checkmark$
 \\ \hline
 $N_{6x}$ & $\times$ & $\times$ & $\times$ & $\checkmark$ &  $\checkmark$  & $\checkmark$ & $\times$ & $\times$ &$\checkmark$
 \\ \hline
 $N_{1y}$
        &$\times$ & $\times$ &$\times$ &$\checkmark$  &  $\checkmark$    & $\checkmark$ & $\checkmark$ & $\times$& $\checkmark$
        \\
        \hline
	  $N_{2y}$
	  & $\times$ & $\checkmark$ & $\times$ & $\times$ &  $\checkmark$  & $\checkmark$ & $\times$ & $\times$ &$\checkmark$
	  \\
	  \hline
	  $N_{3y}$ & $\times$ & $\checkmark$ & $\times$ & $\times$ &  $\checkmark$  & $\checkmark$ & $\times$ & $\times$ &$\checkmark$
	  \\
	  \hline
	  $N_{4y}$ & $\times$ & $\times$ &$\times$ &$\checkmark$  &  $\checkmark$    & $\checkmark$ & $\checkmark$ & $\times$& $\checkmark$
 \\
 \hline
	  $N_{5y}$ & $\times$ & $\checkmark$ & $\times$ & $\times$ &  $\checkmark$  & $\checkmark$ & $\times$ & $\times$ &$\checkmark$\\
		\hline\hline
	\end{tabular*}
\end{table*}
\bibliography{bib} 
\bibliographystyle{unsrtnat}
\end{document}